# FKreg: A MATLAB toolbox for fast Multivariate Kernel Regression

Ying Wang [a], Min Li [a], Deirel Paz-Linares [a, b], Maria L. Bringas Vega [a, b], Pedro A. Valdés-Sosa [a, b],*

[a] *The Clinical Hospital of Chengdu Brain Science Institute, MOE Key Lab for Neuroinformation, School of Life Science and Technology, University of Electronic Science and Technology of China, Chengdu, China*
[b] *Cuban Center for Neuroscience, La Habana, Cuba*

## Abstract

Kernel smooth is the most fundamental technique for data density and regression estimation. However, time-consuming is the biggest obstacle for the application that the direct evaluation of kernel smooth for $N$ samples needs $O(N^2)$ operations. People have developed fast smooth algorithms using the idea of binning with FFT. Unfortunately, the accuracy is not controllable, and the implementation for multivariable and its bandwidth selection for the fast method is not available. Hence, we introduce a new MATLAB toolbox for fast multivariate kernel regression with the idea of non-uniform FFT (NUFFT), which implemented the algorithm for $M$ gridding points with $O(N + M \log M)$ complexity and accuracy controllability. The bandwidth selection problem utilizes the Fast Monte-Carlo algorithm to estimate the degree of freedom (DF), saving enormous cross-validation time even better when data share the same grid space for multiple regression. Up to now, this is the first toolbox for fast-binning high-dimensional kernel regression. Moreover, the estimation for local polynomial regression, the conditional variance for the heteroscedastic model, and the complex-valued datasets are also implemented in this toolbox. The performance is demonstrated with simulations and an application on the quantitive EEG.

## Highlights

- We implemented the fast kernel regression in pure MATLAB code for a large sample size problem using type-1 NUFFT, which improved accuracy and speed.
- We extended this tool for multivariate local polynomial regression.
- In order to adapt to the fast algorithm bandwidth selection, this tool uses the fast Monte-Carlo algorithm to estimate the DF with a few trials of Gaussian noise to avoid the conventional method with expensive time and space complexity.
- This toolbox includes variance estimation for solving the heteroscedasticity model, and the confidence interval also is supplied.

- This toolbox can solve complex value regression, as in spectrum fitting.

## Introduction

*This paper will denote scalar quantities with lower case letters, e.g. $x$, constant scalar with $\mathrm{X}$; vectors with lower base bold italic letters, e.g. $\boldsymbol{x}$, matrices with uppercase bold italic, e.g. $\boldsymbol{X}$, marked the estimate with a hat $\hat{\boldsymbol{x}}$.*

Smoothing techniques provided a valuable service for time series and graphic processing. Suppose $(\boldsymbol{X}, \boldsymbol{y}) = \{(\boldsymbol{x}_1, y_1), (\boldsymbol{x}_2, y_2), \ldots (\boldsymbol{x}_N, y_N)\}$ a random sample scatter from a strictly stationary process, which is independent and identical, where $\boldsymbol{x}_i \in \mathbb{R}^d$ have a common density function $f$, and $y_i$ is a scalar response. Let $m(\boldsymbol{x}) = E(y | \boldsymbol{x}_i = \boldsymbol{x})$ is the smooth function and $\sigma^2(\boldsymbol{x}) = \mathrm{Var}(y | \boldsymbol{x}_i = \boldsymbol{x})$ is the conditional variance.

The noise response $\boldsymbol{y}_i$ be modeled as follows,

$$y_i = m(\boldsymbol{x}_i) + \sigma(\boldsymbol{x}_i)\varepsilon_i, \quad i = 1, \ldots, N \tag{1}$$

Where $\varepsilon_i$ is independent and identically distributed (i.i.d.) that, $E(\varepsilon_i | \boldsymbol{x}_i) = 0$ and $\mathrm{Var}(\varepsilon_i | \boldsymbol{x}_i) = 1$. Popular methods for estimate $m(\cdot)$ include the Nadaraya-Watson (NW) kernel estimator (Nadaraya, 1964; Watson, 1964), the Gasser-Muller kernel estimator (Gasser et al., 1984), the local weighted polynomial regression (LWP), and the spline smoother(Schoenberg, 1964; Wahba, 1990).

Kernel smooth without restrictions of parametric models and understanding intuitively has been a powerful tool to reveal data structure and development trends (W. Härdle et al., 2004; Loader, 1999; Witten et al., 2009). However, the NW kernel and Gasser-Muller have flaws that at the boundary of prediction space, the kernel space is asymmetric, and the estimator has bias caused by the slope, which has been discussed by(Chu and Marron, 1991). The improved estimator local polynomial regression reduces the boundary bias efficiently (Fan, 1991). For example, the least square estimator for local polynomial $m(\boldsymbol{x}_i)$

$$\text{minimize} \sum_{i=1}^{N} \left\{ y_i - \sum_{l=0}^{\ell} \boldsymbol{\beta}_l^T (\boldsymbol{x}_i - \boldsymbol{x})^l \right\}^2 K_h(\boldsymbol{x}_i - \boldsymbol{x}) \tag{2}$$

where $\ell$ is the order of a polynomial. $K_h$ is the scaled kernel function, $K_h(\boldsymbol{u}) = |\boldsymbol{h}|^{-1/2} K(\boldsymbol{h}^{-1/2}\boldsymbol{u})$ where $K$ is unscaled kernel function which could kernel functions like Gaussian, uniform, bi-weight, tri-weight, and Epanechnikov et al. The solution of equation (2) is

$$\hat{m}(\boldsymbol{x}) = \boldsymbol{e}_1 \left( \boldsymbol{D}_{\ell,\boldsymbol{x}}^T \boldsymbol{W}_{\boldsymbol{x}} \boldsymbol{D}_{\ell,\boldsymbol{x}} \right)^{-1} \boldsymbol{D}_{\ell,\boldsymbol{x}}^T \boldsymbol{W}_{\boldsymbol{x}} \boldsymbol{y} \tag{3}$$

where $\boldsymbol{e}_1 = [1, \ldots, 0]^T$. $\boldsymbol{e}_i = [0, \ldots, 1, \ldots, 0]_i \in \mathbb{R}^d$ whose $i$–th element is 1 and the rest are zeros, specifically $\boldsymbol{e}_0 = [0, \ldots, 0, \ldots, 0]$

$$\boldsymbol{D}_x = \begin{bmatrix} 1 & (\boldsymbol{x}_1 - \boldsymbol{x})^T & \text{vech}\left[(\boldsymbol{x}_1 - \boldsymbol{x})(\boldsymbol{x}_1 - \boldsymbol{x})^T\right] & \cdots \\ \vdots & \vdots & \vdots & \vdots \\ 1 & (\boldsymbol{x}_N - \boldsymbol{x})^T & \text{vech}\left[(\boldsymbol{x}_N - \boldsymbol{x})(\boldsymbol{x}_N - \boldsymbol{x})^T\right] & \cdots \end{bmatrix} \qquad (4)$$

where $\boldsymbol{W}(\boldsymbol{x})$ is the $N \times N$ symmetric positive defined weight matrix with the kernel density as diagonals.

$$\boldsymbol{W}(\boldsymbol{x}) = diag\left([K_h(\boldsymbol{x}_i - \boldsymbol{x})]\right), \; i = 1:N \qquad (5)$$

An equation (3) could also be rewritten as $\hat{m}(\boldsymbol{x}) = \boldsymbol{e}_1 \boldsymbol{S} \boldsymbol{y}$, where $\boldsymbol{S} == \left(\boldsymbol{D}_x^T \boldsymbol{W} \boldsymbol{D}_x\right)^{-1} \boldsymbol{D}_x^T \boldsymbol{W}$ is the smooth matrix.

The smooth function of regression can lie on an equal space

$$g_{1,j} < \ldots < g_{k_j,j} < \ldots < g_{M_j,j}, \; 1 \leq j \leq d \qquad (6)$$

$M_j$ is the number of grides in $j-$th direction. The core of kernel regression is the local weight which depends on the kernel calculation, the primary time-consuming and storage spending process involves $O(NM)$ calculations, where $M = M_1 \times M_2 \times \ldots \times M_d$. With the increasing sample size, the computing effort of direct estimation goes up drastically, especially for high dimensional regression problems and even worse with multi-response. Therefore, a fast and stable algorithm for the computation of the estimated surfaces is desperately needed. For example, the "updating method" and "binning method" are two strategies for efficient evaluation.

The updating method avoids the redundant summation operations of convolution. The recent study (Langrené and Warin, 2019) improved the stability of the "updating method. The bandwidth selection is still a problem since it has to recalculate the whole process for different bandwidths. Although updating method offers adaptive bandwidth, the result in the statistic sense is inefficient, and interpolant is hard (Fan and Marron, 1994).

By gridding data into equal space with the desired number of gridding points, the number of bins is usually smaller than the number of the points of input samples. The idea of the binning method is to summarize the samples on the uniform binning points (W. K. Härdle and Scott, 1992; Silverman, 1986) with $O(N)$ (Fan and Marron, 1994). Then, the uniform grid space allows the usage of Fast Fourier Transform (FFT) $O(M \log M)$, which greatly reduces the evaluations number of kernel estimations to $O(M)$ (Cleveland and Grosse, 1991; Fan and Marron, 1994; Wand, 1994). For regression function estimation or scatterplot smoother, the goal is usually to get the whole smooth function, later estimating points in range. Thus, the binning method for smooth is a much more rational choice. However, this sped up the calculation at the expense of accuracy.

From the point of view of signal processing is more intuitive to introduce the Fourier transform. The kernel regression is equivalent to a low pass filter, especially when the kernel is single side, i.e., the kernel $K_h = 0$ when $x < 0$, the regressor is equal to a causal low pass filter. Therefore, the technique in signal processing can be introduced, the regressor can also be studied with frequency domain property (Schlax and Chelton, 1992), and the good property of local polynomial can also be used for filter design (Proietti and Luati, 2008, 2011; Zeng et al., 2009). This paper uses the non-uniform sampled signal processing method NUFFT (non-uniform FFT) (Dutt and Rokhlin, 1993) for gridding and transformation in the binning method. The advantage is that the accuracy compared with direct FFT, which corresponds to the direct method of kernel regression, is controllable.

Additionally, even the algorithm of fast binning regression has been studied a lot. However, several legacy problems still have not been solved efficaciously.

- Degree of Freedom (DF) estimation for model selection criteria calculation;
- Empirical bandwidth estimation or selection based on cross-validation (CV) calculation;
- Heteroscedasticity error variance estimation;
- Complex data regression includes the complex observation with real locations and the complex observation with complex locations.

The essential parameter for DF and CV evaluation requires estimating the smooth matrix $S$ where $\hat{m}(x_i) = e_1 S y$. Following the same idea with ordinary kernel regression, Turlach and Wand (1996) instead of working on the actual data point, move to the binned grids that treat $S$ as a function that maps the grid count to the binned smoother on grid points and involved the same calculation with fast kernel estimation. However, this fast kernel bandwidth selection is redundant when the multiple observations or measurements lie on a common original space, known as the multi-response problem. Additionally, the previous study about fast kernel regression only talked about the homoscedastic model for the property of error variance.

In this paper, we provide the general fast polynomial regression code which made several improvements as follows,

- Extend the regression from univariate to multivariate and support local polynomial regression;
- Efficient estimation of multiple multivariate local polynomial regression. Here, with the symbolic calculation, the expression of the local polynomial can be easily achieved, and the data can be assigned to it and calculated with vectorization efficiency. Instead of calculating responses one by one, we evaluate multiple regressions simultaneously since they share the same smooth range;

- Speed-up bandwidth selection. We use the Fast Monte Carlos to estimate DF. Each trial needs the same amount of fast regression but is still much more efficient than the convention DF estimation. In addition, the fast Monte Carlos is independent of responses. Once the prediction range is defined, the DF will be confirmed. Thus, for multi-response that share the same alternative bandwidth range, the DF only needs to be calculated once, which reduces computation intensively;
- Heteroscedasticity variance estimation.
- Compatible with complex value regression.

The structure of this paper is following. We first give the methodology illustrations of multiple multivariate fast kernel regression in section 1. Section 2 talks about the fast bandwidth selection. Section 3 gives the expression of the conditional variance estimation based on fast kernel regression. Section 4 introduces the management of the FKreg toolbox with examples with corresponding conventions and the application of complex-value data. The speed comparison of this toolbox with the fast kernel regression toolbox adopts the fast sum updating method in section 5.

# 1 Fast multivariable kernel regression for Non-uniform data

The kernel regression holds the same idea as the filter(Ouassou et al., 2016), which convolves the discrete samples with a kernel function. On the uniform sampled data, they can be accelerated by the FFT. Here we first introduce the fast multivariable kernel regression with uniform sampling, then discuss the use fast gaussian gridding method to convert the non-uniform sampled data to uniform.

## 1.1 Fast multivariable kernel method

The Nadaraya-Watson kernel regression is a particular estimator of the equation (3), which can be written as,

$$\hat{m}(\boldsymbol{x};0) = \frac{\sum_{i=1}^{N} K_h(\boldsymbol{x}_i - \boldsymbol{x}) y_i}{\sum_{i=1}^{N} K_h(\boldsymbol{x}_i - \boldsymbol{x})} \quad (7)$$

where the denominator is known as the kernel density estimation (KDE),

$$\hat{f}(\boldsymbol{x}) = \frac{1}{n} \sum_{i=1}^{N} K_h(\boldsymbol{x}_i - \boldsymbol{x}) \quad (8)$$

which can be considered as the convolution of the continuous kernel function with the Dirac delta function on sample location. Flowing the convention of (Fan and Marron, 1994), let

$$\hat{s}_\ell(\boldsymbol{x}) = \sum_{i=1}^{N} K_h(\boldsymbol{x} - \boldsymbol{x}_i)(\boldsymbol{x} - \boldsymbol{x}_i)^\ell$$

$$\hat{t}_\ell(\boldsymbol{x}) = \sum_{i=1}^{N} K_h(\boldsymbol{x} - \boldsymbol{x}_i)(\boldsymbol{x} - \boldsymbol{x}_i)^\ell y_i \quad (9)$$

where $\ell$ is the order of the model. Thus, the multivariate NW kernel regression can be written as the flow with $\ell = 0$,

$$\hat{f}(x) = \frac{1}{n}\hat{s}_0(x)$$
$$\hat{m}(x;0) = \frac{\hat{t}_0(x)}{\hat{s}_0(x)} \tag{10}$$

Generalize the local polynomial estimator for multivariate with similar but simpler expressions under the uniform sampling condition (Wand, 1994). Consider the equal space grid in each direction as in equation (6), and the equal space gap at direction $j$ denotes with $\tau_j = (g_{M_j,j} - g_{M_1,j})/(M_j - 1)$. Let $g_k = (g_{k_1,1}, g_{k_2,1}, \ldots, g_{k_d,d})$, $1 \leq k_j \leq M_j$, $j = 1,\ldots,d$ denote the grid point with index $k = (k_1,\ldots,k_d)$.

The fast-binning methods of kernel estimation work on the gridded value $u$ of the Dirac delta function on sample location $\delta(x-X)$ and the gridded value $v$ of the sampled response $y$, which means distributing the amount of original data $(X, \delta(x-X))$ and $(X, y)$ into grid points. Section 1.2 will give the acquisition strategy of fast griding. The binned approximation of (3) is,

$$\hat{m}(g_k;\ell) = e_1^T \hat{S}_{\ell,g_k}^{-1} \hat{t}_{\ell,g_k} \tag{11}$$

where the $\hat{S}_{\ell,g_k}$ and $\hat{t}_{\ell,g_k}$ has high order terms differ with $\hat{s}_0(x)$ and $\hat{t}_0(x)$

$$\hat{S}_{\ell,g_k} = \sum_{k_d=1}^{M_d} \cdots \sum_{k_1=1}^{M_1} d_{\ell,g_{k'}-g_k}^T w_{g_{k'}-g_k} d_{\ell,g_{k'}-g_k} u_{g_{k'}} \tag{12}$$

$$\hat{t}_{\ell,g_k} = \sum_{k_d=1}^{M_d} \cdots \sum_{k_1=1}^{M_1} d_{g_{k'}-g_k}^T w_{g_{k'}-g_k} v_{g_{k'}} \tag{13}$$

Since the $\hat{S}_{\ell,g_k}$ and $\hat{t}_{\ell,g_k}$ are calculated by the convolution on the uniform data, the calculation can accelerate by the FFT. In the implementation, the matrix operation of the design matrix $d_{\ell,x}$ and the kernel weighted matrix $w_x$ can be computed with symbolic calculation and then assigned the grid points into it.

$$d_x = \begin{bmatrix} 1 & x^T & \text{vech}(x^T x) & \cdots \end{bmatrix} \tag{14}$$
$$w_x = K_h(x) \tag{15}$$

The convolution for each element of $\hat{S}_{\ell,g_k}$ and $\hat{t}_{\ell,g_k}$ with the grided value can be efficiently done in the frequency domain.

After the kernel was evaluated and transformed back to the original space, the uniform sampled $\hat{S}_{\ell,g_k}$ and $\hat{t}_{\ell,g_k}$ will be used to calculate $\hat{m}(g_k;\ell)$. Instead of using type-2 NUFFT, uniform to non-uniform transformation, to predict, we use the efficient and flexible interpolation implementation *griddedinterpolant* class in MATLAB. The interpolation and extrapolation methods can be customized.

## 1.2 Non-uniform Fast Fourier Transform

Gridding can also be considered as a convolution operation, although the kernel weights for gridding in some binning methods may vary with location.

$$f_\tau(x_j) = \sum_{i=1}^{N} w_{i,j} f(x_i) \tag{16}$$

where the $f_\tau(\cdot)$ is the gridded value of the non-uniform sampled data $f(\cdot)$. The gridding of binning methods in previous studies sacrificed the accuracy to achieve fast computation.

In order to achieve higher accuracy of gridding, here we introduce the NUFFT. In this paper, we use the type-1 NUFFT transform non-uniform data to spectral domain with uniform sampling, which gathers the gridding and the FFT transform as discussed above. NUFFT is wildly used for Fourier transformation for the non-uniform sampled data (Anderson and Dahleh, 1996; Bagchi and Mitra, 2012; Dutt and Rokhlin, 1993; Greengard and Lee, 2004). The NUFFT with fast gaussian gridding used in this paper applies the gridding kernel with a global bandwidth which can be deconvolved in the frequency domain to compensate for its blurring effect.

The fast Gaussian gridding trusses Gaussian kernels for points at each direction in the original space and smears the weighted to nearby grid points as the target position to be assigned. Give the illustration with a 1-d dataset. Consider signal $f(x)$ is sampled non-equidistant in $[0, 2\pi]$. To achieve the requirement of uniformity of standard FFT, we can get the uniform located response of the signal through a broad bandwidth kernel. The gridded value can be evaluated as flow,

$$f_\tau(2\pi m / M_r) = \sum_{i=0}^{N-1} f_i(x) \sum_{l=-\infty}^{\infty} e^{-(x_i - 2\pi m/M_r - 2l\pi)^2 / 4\tau} \tag{17}$$

where $M_r$ is the length of FFT. $M_r = 2L$ is suggested in Greengard and Lee(2004), which means the data will be upsampled with this sharp Gaussian kernel. Specifically, to achieve efficiency for extremely large sample size cases, since the sample is dense enough, the error between binning and direct regression methods can be tolerated, and the gridding number can be set to $M_r = L$. On all accounts, since the Gaussian function is sharp, it only needs to consider nearby points. Therefore, the convolution result $f_\tau$ can be achieved efficiently with $O(N)$.

Remarkably, bandwidth in signal processing differs from kernel bandwidth in kernel regression. The broad bandwidth in the signal processing means small bandwidth in kernel regression, which keeps more high-frequency components. It corresponds to being more unsmooth for the regression result or says it takes more account of local variation and is more sensitive to the noise.

To avoid the wrap-around effect of circular convolution, we padded zeros before the FFT. The length of the zeros can be set with the nextpow2

$$\tilde{f}_\tau = \begin{bmatrix} \mathbf{0} & f_\tau & \mathbf{0} \end{bmatrix}^T \tag{18}$$

The boundary effect of padding zeros will be counteracted between the *u* and *v* in formula (11). After the convolution is evaluated, the FFT will be applied

$$F_\tau(k) \approx \frac{1}{M_r} \sum_{m=0}^{M_r - 1} \tilde{f}_\tau(2\pi m / M_r) e^{-ik 2\pi m / M_r} \tag{19}$$

As mentioned, to have higher accuracy, the effect of this sharp gridding kernel will be deconvolved from the signal in the frequency domain,

$$F(k) = \sqrt{\frac{\pi}{\tau}} e^{k^2 \tau} F_\tau(k) \tag{20}$$

The $F(k)$ has very high accuracy compared with typical DFT. The accuracy in digits is controlled by the number $M_{sp}$ of nearest spreading points of each source point (Dutt and Rokhlin, 1993; Greengard and Lee, 2004).

Finally, the gridded value we want can be directly obtained by the IFFT (inverse-FFT).

$$f(x_j) = \sum_{k=-\frac{M}{2}}^{\frac{M}{2}-1} F(k) e^{ikx_j} \tag{21}$$

With this technique, the kernel density and $y$ can be gridded into the uniform grid. The complexity of this procedure depends on the gridding step $O(N)$ and the FFT $O(M \log M)$. Kernel regression always uses the $M \ll N$ is to reduce the further computation(Fan and Marron, 1994), which is also used in other nonparametric methods (Snelson and Ghahramani, 2005).

## 2  Bandwidth selection

### 2.1 Fast Monte-Carlo estimation of DF

The criteria of bandwidth selection require a set of cross-validation, such as least square cross-validation (LSCV) (Stone, 1984), bias cross-validation (BCV) (Sain et al., 2012), et al. And all of the calculation of CV value needs the approximation degree of freedom (DF) of data. For example, the generalized cross-validation (GCV),

$$\text{GCV}_h = \frac{\frac{1}{N} \|y - S_h y\|^2}{\left[\frac{1}{N} tr(I - S_h)\right]^2} \tag{22}$$

The numerator is estimated square error, and the denominator contains DF (the trace of the smooth matrix). As mentioned before, the fast bandwidth selection is an uneconomical way to multi-response. Girard (1989) used a fast Monte-Carlo algorithm to calculate the reliable estimates of DF. The algorithm is as follows:

  i. Generate random vector $\mathbf{w}_p$, $p = 1, \ldots, Np$, that $\mathbf{w}_i \sim \mathcal{N}(0, \mathbf{I}_N)$.

  ii. Estimate $\hat{\mathbf{w}}_p^{(h_k)}$ at bandwidth $h_k$, using the fast kernel regression method in section 1 for each bandwidth in the $[h_1, \ldots, h_k, \ldots, h_{Nh}]$;

iii. Approximate the mean value of $(1/N)\operatorname{tr}(\boldsymbol{I}-\boldsymbol{S}_{h_k})$ with $(1/N)\mathbf{w}_p^T(\mathbf{w}_p-\hat{\mathbf{w}}_p^{(h_k)})/(\mathbf{w}_p^T\mathbf{w}_p)$ and obtain the DF value $\operatorname{tr}(\boldsymbol{S}_{h_k})$ at the bandwidth $h_k$.

iv. According to the equation (22), the estimate of $\text{GCV}_{h_k}$ is calculated with the mean square error $\mathbf{y}$.

To be mentioned, the selection of $Np$ will not be required more than 20, which be testified in Girard (1989) and can also be proved in the following sub-section.

## 2.2 Verification of Fast Monte-Carlo estimation

To test the reliability of the Fast Monte-Carlo (FMC) estimation, we compare the estimates of $(1/N)\operatorname{tr}(\boldsymbol{I}-\boldsymbol{S}_h)$, $\operatorname{tr}(\boldsymbol{S}_h)$, DF and GCV with the FMC algorithm and standard NW calculation. In the FMC calculation, we used the fast NW regression and set $Np=10$ and $Np=30$ to determine the effect of iteration numbers for FMC estimation. The verification for other smooth methods like polynomial spline and thin-plat spline in was discussed in Girard (1989).

Here, we use the $\operatorname{sinc}$ function with small noise as an example that $\text{yr}=\operatorname{sinc}(x)+\varepsilon$ with $x=[-100:0.2:100]$. Figure 1 shows the regression of $\text{yr}$ three estimates. First, these three estimations perform very well with the automatic bandwidth selection. Furthermore, when zooming into detail, the FMC was estimated with $Np=30$ closer to ordinary NW estimation than $Np=10$, which, as expected, shows the larger $Np$ gives more reliable results than the small $Np$.

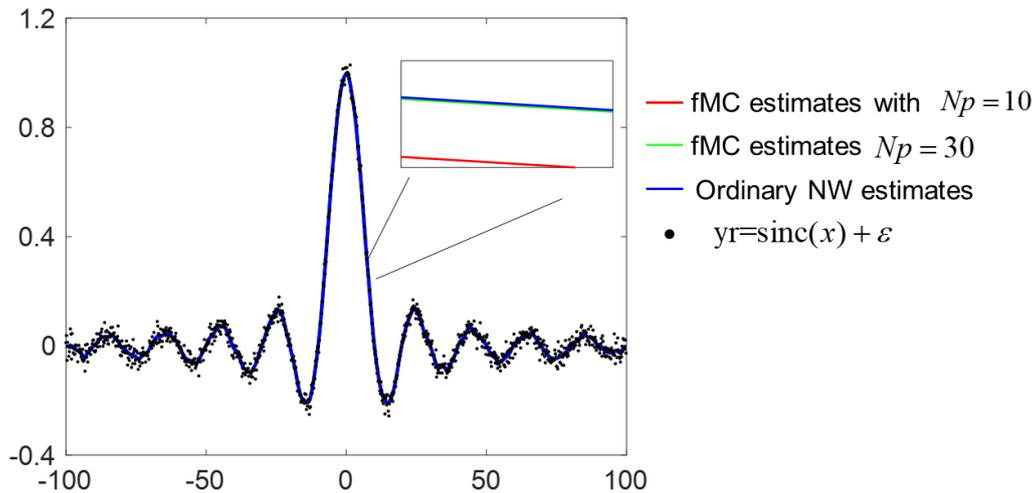

*Figure 1: univariate NW regression with the ordinary and fast algorithm. The red and green lines are fast NW regression with Np equal to 10 and 30 with optimal bandwidth selected by rGCV. The blue line is ordinary NW regression in which bandwidth is selected by GCV. The black dots are 1000 real points from -100 to 100.*

We compared the values of $(1/N)\operatorname{tr}(\boldsymbol{I}-\boldsymbol{S}_h)$, $\operatorname{tr}(\boldsymbol{S}_h)$, MSE and GCV shown in Figure 2. There is no evident difference between these three estimates. The selection of optimal bandwidth of FMC with $Np=30$ is the same as the ordinary NW that $h_{opt}=0.73$ and it is very close to the optimal bandwidth of FMC with $Np=10$, that $h_{opt}=0.77$. The minimum difference of DF between these three estimates under their optimal bandwidth is 0.93 (ordinary with 10 iterations for the FMC estimate). The difference of GCV values at the level of 1e-6.

The results show that the FMC algorithm for DF estimation is reliable, and the iteration number $Np=10$ is enough. Girard's paper also mentioned that the iteration number would never be larger than 20.

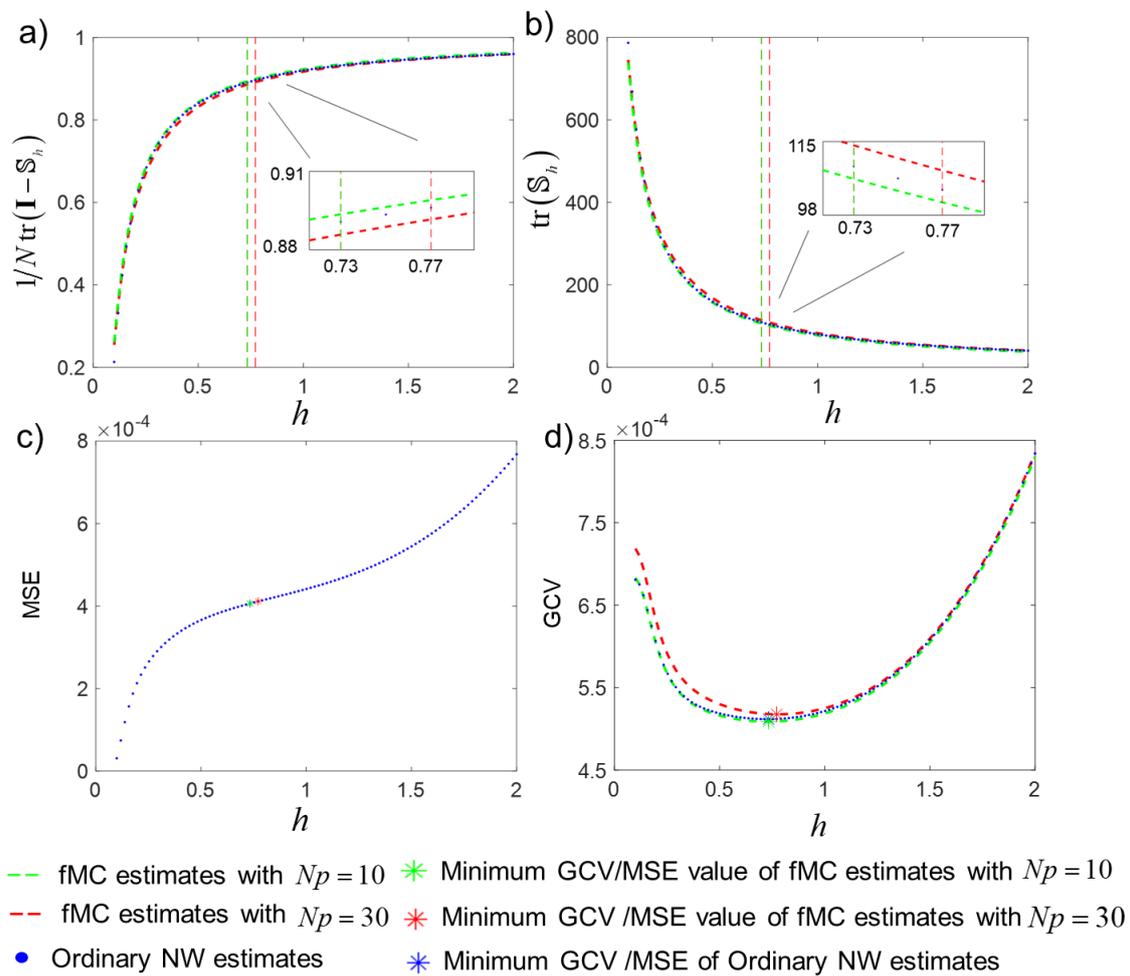

—— fMC estimates with $Np=10$  ✳ Minimum GCV/MSE value of fMC estimates with $Np=10$
—— fMC estimates with $Np=30$  ✳ Minimum GCV/MSE value of fMC estimates with $Np=30$
● Ordinary NW estimates          ✳ Minimum GCV/MSE of Ordinary NW estimates

Figure 2: Performance of GCV values. a) the values of $(1/N)\operatorname{tr}(\boldsymbol{I}-\boldsymbol{S}_h)$ which are close to 1 when the bandwidth increase; b) the values of df, which go to zero when the bandwidth becomes larger; c) the MSE of estimates. d) GCV values. A local cone of GCV values is used to select the optimal bandwidth. The green dotted lines are performances of fast kernel regression with $Np=10$. The red dotted lines correspond to kernel regression with $Np=30$ and the blue dots are the results of the ordinary kernel method.

## 3   Variance estimation

Directly applying the kernel regression to the heteroscedasticity data cannot get the proper surface trend because of the unsymmetric distribution of variance. For heavy-tailed innovation, Chen et al., (2009) developed a variance reduction technique to improve the inference for conditional variance.

Recall equation (1), the local linear estimation obtained by minimizing the cost function of (2). Similarly, the estimate of $\sigma^2(x)$ is $\hat{\sigma}^2(x) = \hat{\alpha}_1$ with the cost function here we demonstrate with a univariate case

$$(\hat{\alpha}_1, \hat{\alpha}_2) = \mathrm{argmin}_{(\alpha_1,\alpha_2)} \sum_{i=1}^{n} \{\hat{r}_i - \alpha_1 - \alpha_2(x_i - x)\}^2 W\left(\frac{x_i - x}{h_\alpha}\right) \qquad (23)$$

where $\hat{r}_i = \{y_i - \hat{m}(x_i)\}^2$

There is an essential problem that this conditional variance estimator is not always positive. The reasonable strategy is using $\log \sigma^2(x)$ instead of $\sigma^2(x)$, for example, the local Normal likelihood estimation (FAN and YAO, 1998; Yu and Jones, 2012). This ensures that the conditional variance estimator always is positive. Chen et al., (2009) by considering the empirical situation of the existence of heavy tails data, proposed a new estimation for $\sigma^2(x)$ that with the equation

$$\log r_i = v(x_i) + \log(\varepsilon_i^2 / \kappa) \qquad (24)$$

where $r_i = \{y_i - m(x_i)\}^2$, $v(x) = \log(\kappa \sigma^2(x))$ with constrain that $E\{\log(\varepsilon_i^2 / \kappa)\} = 0$. First, the estimate of $v(x)$ is $\hat{v}(x) = \hat{b}_1$

$$(\hat{b}_1, \hat{b}_2) = \mathrm{argmin}_{(b_1,b_2)} \sum_{i=1}^{n} \{\log(\hat{r}_i + N^{-1}) - b_1 - b_2(x_i - x)\}^2 W\left(\frac{x_i - x}{h}\right) \qquad (25)$$

here, to avoid $\log(0)$, employ $\log(\hat{r}_i + n^{-1})$ instead of $\log(\hat{r}_i)$. Since $E\{\log(\varepsilon_i^2 / x_i)\} = 0$ and $r_i = \exp\{v(x_i)\}\varepsilon_i^2 / \kappa$, the $\kappa$ can be estimated by $\hat{\kappa} = \left[\frac{1}{N}\sum_{i=1}^{N} \hat{r}_i \exp\{-\hat{v}(x_i)\}\right]^{-1}$, so the estimator of $\sigma^2(x)$ will be defined as,

$$\hat{\sigma}^2(x) = \exp\{\hat{v}(x)\} / \hat{\kappa} \qquad (26)$$

## 4   Toolbox and experiments

The toolbox is implemented with pure MATLAB code. The operation illustrated above is arranged to avoid redundant calculations in our FKreg toolbox.

Here is a brief overview of the usage of the toolbox. Figure 3 shows the steps of the whole workflow and the definition of parameters shown in Table I.

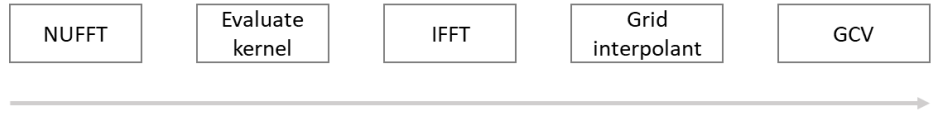

*Figure 3: The algorithm workflow of the fast kernel regression toolbox.*

The function **cv_fkreg** is the primary entrance of the toolbox **[regs]=cv_fkreg(x,yr,hlist,opt),** that x and yr are pairwise datasets, *hlist* is the bandwidth candidates, and opt is a structure including all the regression parameters. The function **cv_fkreg** includes essential components inside as follows:

*fkreg_create:* Initial parameters in variable opt.

Notice that the number of grid points in each direction in the toolbox does not have to be equal. The default value for each direction is set to the same value $M_j = R\left\lceil \sqrt[d]{N} \right\rceil$. The number of zero-padded grid points is $L_j = \text{nextpow2}(M_j)$ where $\text{nextpow2}(x) = 2^{\lceil \log_2(x) \rceil}$. The number of points for the FFT algorithm is $Mr_j = 2L_j$. Despite initializing the number of points, this function will also generate the kernel function on the uniform grid, including high order kernel for local polynomial regression and the symbolic result of the formula (11).

*Table 1: Setting of main parameters of FKreg toolbox.*

| Parameter | Definition | Default setup |
|---|---|---|
| y_type | A string choose the type of response, it can be 'mean' or 'variance' | 'Mean' |
| kernel_type | A string sets the kernel type for fast kernel regression. The kernel type provided here is gaussian, Epanechnikov box, triangle, etc. | 'gaussian' |
| order | A scalar sets the order of local polynomial regression | 0 |
| R | A scalar sets the gridding ratio for the number of grid points in each direction $M_j = R\lceil \sqrt[d]{N} \rceil$. | 1 |
| M | A scalar or a vector sets grid points $M_j$ in each direction. It has a higher priority than opt.R | $M_j = R\lceil \sqrt[d]{N} \rceil$ |
| flag_power2 | Bool value sets if padding data with zeros to avoid boundary effect. | true |
| accuracy | A scalar sets accuracy for the NUFFT, i.e., the number of nearest points used in fast gaussian gridding. | $max(6-\lceil log10(Tx) \rceil, 1)$ |
| nufft_deconv | Bool value sets if do the deconvolution for the gridding kernel. | accuracy>6 |
| dstd | A scalar sets the GCV standard deviation, which use to obtain the $h_{opt}$ with more relaxed GCV value, $h_{optimal} = \max\{h \mid GCV(h) < (GCV(h) + \text{dstd} \cdot (\text{std}(GCV(h))))\}$ | 0 |
| calc_dof | Bool value sets if run the bandwidth selection or save results for all bandwidth. | true |
| Np | A scalar sets the iteration times of FMC estimation. | 10 |
| y_grid_opt.Method | A string sets the interpolation method for the prediction. It can be changed after the regression. | 'spline' |
| y_grid_opt.ExtrapolationMethod | The string sets the extrapolation method for the prediction can be changed after the regression. | 'linear' |
| compact | Bool value sets if remove non-necessary variables for prediction from the structure. | true |

***fkreg_s***: It applies the NUFFT based fast convolution to calculate the $\hat{s}_\ell$. In the NW regression calculates the density of $x$, i.e., convolve the kernel function with Dirac delta sampling function. The NUFFT code is based on the implementation of Ferrara, (2009) and Vanderplas, (2015)

***fkreg_dof***: Degree of freedom approximation with Fast Monte-Carlo algorithm. This procedure performs the calculation in section 2.1, approximating the value of $(1/n)tr(\mathbf{I} - \mathbb{S}_\tau)$ which will be used in bandwidth selection for GCV calculation and DF $tr(\mathbb{S}_\tau)$ for model selection. The iteration number for this randomized estimation set to 10, is based on the simulation result in section 2.2.

***fkreg_y***: It estimates $\hat{t}_\ell$, $\hat{m}_\ell$ and the bandwidth selection. A task division is used here because running for multi-responses might exhaust memory. The optimal bandwidth can be selected based on the minimal GCV value.

***fkreg_compact***: It selects necessary contents for the prediction.

The input of ***cv_fkreg*** includes the sample points $\mathbf{X}$, observation $\mathbf{y}$, *hlist* (the list of bandwidths, see the function of get_hlist below), and the parameter *opt* structure.

The output of ***cv_fkreg*** is a structure, including $\hat{y}$ and the collection of cross-validation results.

*There are some important auxiliary functions as follows,*

***get_hList***: Generate bandwidth list. The input for 1d regression on log-space, for example is, hlist=get_hList([10],[0.01,1],@logspace) with 10 alternative bandwidths from 0.01 to 1; For bivariate on liner-space is hlist=get_hList([10,20],[0.01,1; 0.1,1],@linspace) that the number and range of alternative bandwidth at two directions can be different. Here in one direction, the number of bandwidth lists is 10 and another is 20 from 0.01 to 1 and 0,1 to 1 separately. The $x$ will be scaled to a standard deviation with 1 during the calculation, so the bandwidth for input is given based on the standardized scale.

**fkreg_predict:** Predicting response on incoming location. Inputs arguments for the function are the structure returned by ***cv_fkreg*** and the new $x$.

***fkreg_interval***: Generating confidence interval. Inputs arguments are the mean and variance estimate by ***cv_fkreg***. The variance could also be a scalar here for the homoscedasticity model.

## 4.1 One-dimensional regression with the example of the Bessel function

We use the Bessel function as an example to present the fast regression of three models with order=0,1,2. We generate 100 bandwidths from 0.01 to 1 in logarithm space and compare the result under their optimal bandwidth for each model, see the code below. The LL and LQ have a good performance at the boundary discussed by Fan and Gijbels, (1992). The higher order it uses, the more high-frequency components will be reserved. In other words, the regressor will be more sensitive to local patterns.

```
rng(0)
N=200; % difine sample size
x=rand(N,1)*20; % define ground truth function
fun=@(x) besselj(1,x); %MATLAB peaks function; %% generate data
y=fun(x);
yr=y+0.1*std(y).*randn(size(y));
hlist=get_hList(100,[0.01,1],@logspace);
opt.order=0; % order of polynomial
[regs]=cv_fkreg(x,yr,hlist,opt);
opt.order=1; % order of polynomial
[regs]=cv_fkreg(x,yr,hlist,opt);
opt.order=2; % order of polynomial
[regs]=cv_fkreg(x,yr,hlist,opt);
```

```
>>
[nw_nufft]: [1d] regression with order 0
 - selected bandwidth - [4th] h
 - normalized scale h =0.046416
 - original scale h =0.26178
Elapsed time is 0.935164 seconds.
[nw_nufft]: [1d] regression with order 1
 - selected bandwidth - [6th] h
 - normalized scale h =0.12915
 - original scale h =0.72843
Elapsed time is 0.141428 seconds.
[nw_nufft]: [2d] regression with order 2
- selected bandwidth - [6th] h
 - normalized scale h =0.12915
 - original scale h =0.72843
Elapsed time is 0.141756 seconds.
```

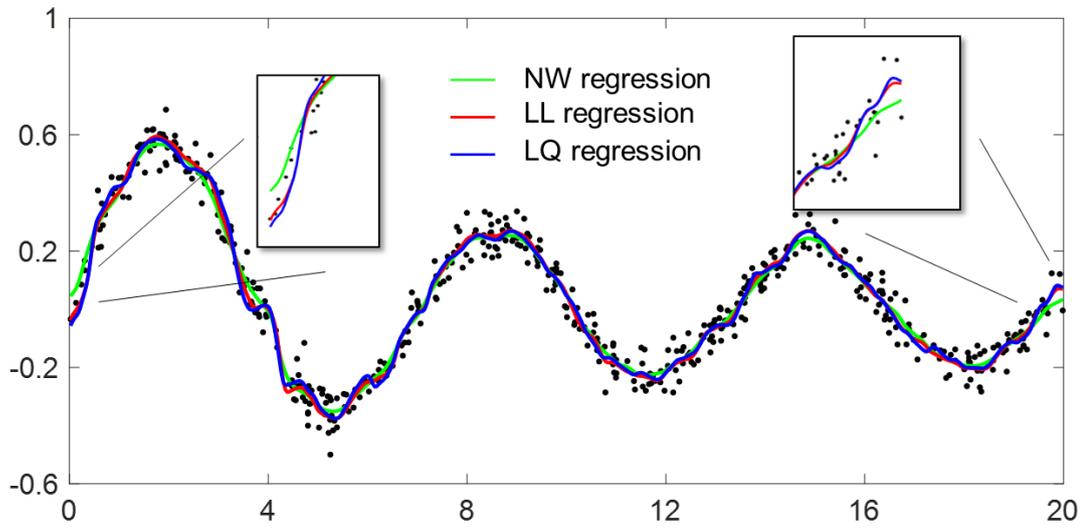

*Figure 4: Univariate fast kernel regression. The green line is the NW regression, the red line is LL, and the blue line is LQ. The LL regression performs better at boundary and describes the shape of the dataset.*

### 4.2 Heavy tailed innovation regression

This section gives examples of heteroscedasticity model regression with the univariate and bivariate datasets. Generate an example dataset with varying variance and estimate the varying variance using the theory in section 3. Consider the kernel regression of the equation (1), the standardized residual is asymptotic to the normal distribution (De Brabanter et al., 2011),

$$\frac{\hat{m}(x) - m(x)}{\sqrt{\Sigma}} \xrightarrow{d} \mathcal{N}(0,1) \qquad (27)$$

where, $\Sigma^2 = \text{var}\left[\hat{m}(x)\right]$ and $\xrightarrow{P}$ indices convergence in distribution. Estimate the $\Sigma$ with FKreg toolbox, the $\hat{\Sigma}$ is convergence to $\Sigma$, probabilitly that $\frac{\hat{\Sigma}}{\Sigma} \xrightarrow{P} 1$. Thus, the pointwise Confidence Interval (CI) approximation with $100(1-\alpha)\%$ for $m(x)$ is,

$$\text{CI} = \hat{m}(x) \pm z_{1-\alpha/2} \sqrt{\hat{\Sigma}}, \qquad (28)$$

The $z_{1-\alpha/2}$ is the $(1-\alpha/2)$-quantile of standard Gaussian distribution.

We give the univariate variance and CI estimate ($\alpha = 0.05$) example with the following MATLAB commands

```
N=10000;
x=2*2*(rand(N,1)-0.5);
fun_mu=@(x) x.^3;
fun_sigma=@(x) 1+4*exp(-(x.^2));
y=fun_mu(x);
s=fun_sigma(x)*0.5*var(y);
yr=y+randn(size(y)).*sqrt(s);
%% mu
opt.order=1;
hlist=get_hList([20],[0.01,1],@logspace);
[regs_mu]=cv_fkreg(x,yr,hlist,opt);
yp=regs_mu.yhat;
%% sigma
opt.order=1;
hlist=get_hList([20],[0.01,1],@logspace);
sr=(yr-yp).^2;
opt.y_type_out='variance';
[regs_sigma]=cv_fkreg(x,sr,hlist,opt);
sp=regs_sigma.yhat;
%% ci
alpha=0.05
ci=fkreg_interval(regs_mu,regs_sigma, alpha);
figure(1)
hold on
plot(x,y,'*r')
scatter(x,yr,'.b','MarkerEdgeAlpha',0.3)
plot_griddedInterpolant(regs_mu.fpp_yhat,[],[],[],LineWidth=2)
plot_interval([],regs_mu,ci)
```

Use the function **fkreg_interval** to calculate the 95% CI based on the estimated $\hat{\Sigma}$ and the function **plot_interval** to plot the CI of $\hat{m}$.

Figure 5-a shows the true $y$ (red stellate), observation $y+\varepsilon$ (blue dots), and the estimated function $\hat{m}$ (solid yellow line). The rose-red dots are simulated variance, which changes with $x$, $\sigma^2 = 1+4\exp(-x^2)$, samples from Gaussian distribution. The variance estimate is very close to $\sigma^2$ in grey line shown in figure 5-b. Figure 5-c shows the corresponding logarithm scale plot. We show the bivariate variance regression and 95% CI estimate in Figure 6.

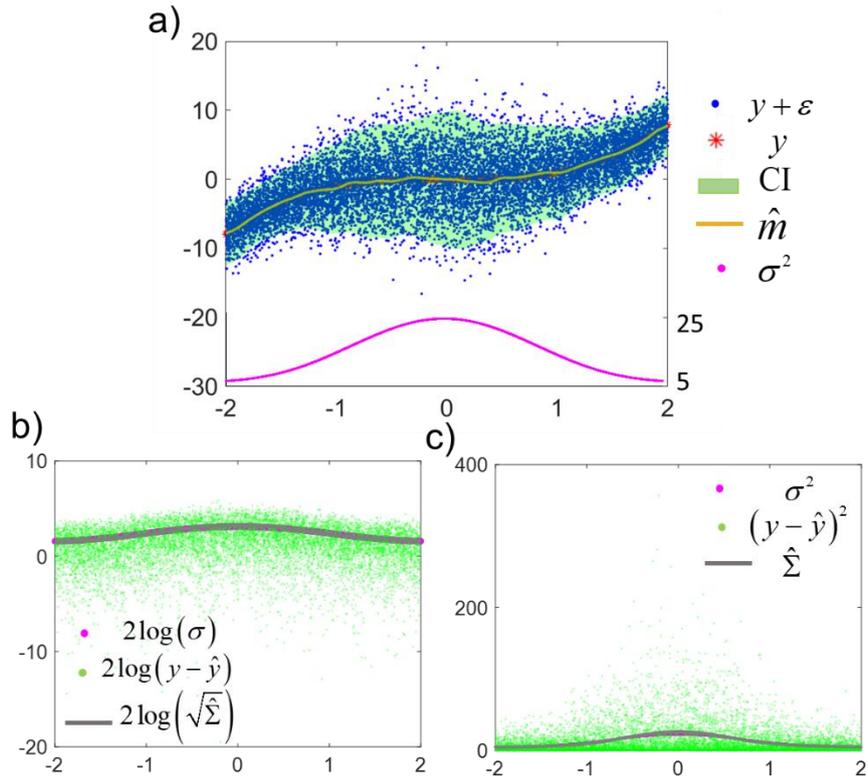

Figure 5: Heteroscedasticity variance estimate of the 1-dimensional dataset. a) The estimates on observation. The red asteroid is the real sampled data $y$, and the blue dots are noisy $y+\varepsilon$. The solid yellow line is the estimation of $y+\varepsilon$ which is overlapped with the red asteroid $y$. The rose-red dots are the variance function for simulation on the sample points, and its estimate is the gray line in subplots b) and c) in different scales where the green dots are sample variance. Because of the non-Gaussian distribution of variance, the presentation of the estimate of variance can be better seen in log-scale in subplot b).

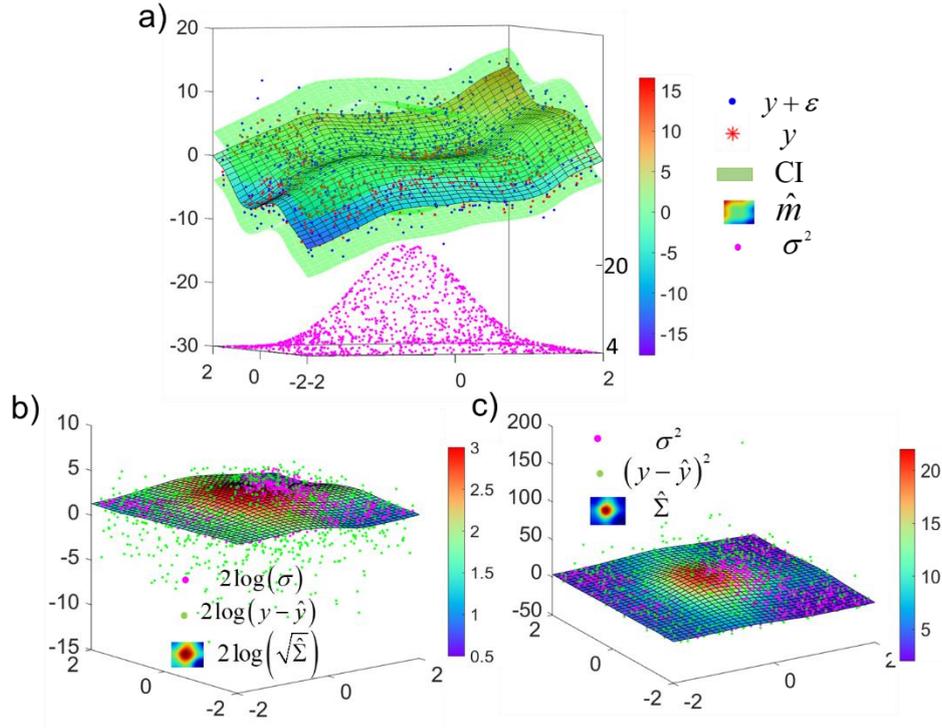

*Figure 6*: Heteroscedasticity variance estimate of the 2-dimensional dataset. a) The estimate of noisy observers. The red asteroid is the real sampled data $y$, and the blue dots are noisy $y+\varepsilon$. The grid surface is the estimate of $y+\varepsilon$ which is overlapped with the red asteroid $y$. The rose-red dots are the variance function on sample position for simulation, and its estimate is grid surface in subplots b) and c) where the green dots are samples of variance. Better viewing of variance estimate in log-scale in subplot b).

### 4.3 Two-dimensional regression for multi-response data

Give a regression example of multi-response here to show the arrangement of multiple regression. We first give an example with simulation to show the usage of commands.

```
N=8100;
x=2*(rand(N,2)-0.5)*3;
fun1=@(x,y) 3*(1-x).^2.*exp(-(x.^2) - (y+1).^2)-10*(x/5 - x.^3 - y.^5).*exp(-x.^2-y.^2)-1/3*exp(-(x+1).^2 - y.^2);
fun2 = @(x,y) (30+(5*x+5).*sin(5*x+5)) .* (4+exp(-(2.5*y+2.5).^2));
fun3 = @(x,y) (cos(x)+cos(y).*exp(-abs(x))+1);
y(:,1)=fun1(x(:,1),x(:,2));
y(:,2)=fun2(x(:,1),x(:,2));
y(:,3)=fun3(x(:,1),x(:,2));
yr=y+0.2*std(y).*randn(size(y));
hlist=get_hList(10,[0.01,0.5],@logspace);
opt.order=1;
tic;[regs]=cv_fkreg(x,yr,hlist,opt);toc
figure
for ifig=1:size(y,2)
    subplot(1,3,ifig)
    hold on;
    opt.iValue=ifig;
    scatter3(x(:,1),x(:,2),y(:,opt.iValue),'.k')
    plot_griddedInterpolant(regs.fpp_yhat,[],[],opt, 'FaceAlpha',0.9);
    set(gca,'FontName','Arial','FontSize',22,'FontWeight','bold');
    colormap(plt.PiYG)
```

```
    view(-50,18)
end
```
The ***cv_fkreg*** can deal with a single data location but several observations simultaneously. Since the data sampled on the same smooth space, therefore, only need one *hlist*. The smooth surface in the griddedInterpolant object can be visualized with the function ***plot_griddedInterpolant*** in the toolbox.

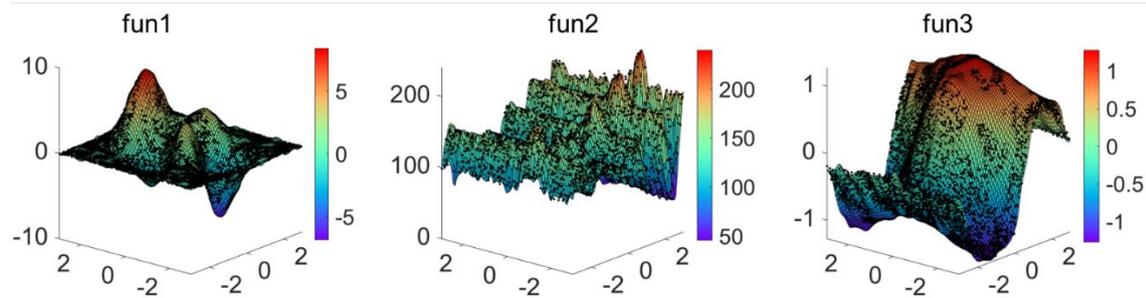

*Figure 7: Multiple 2-dimensional regression. The dots are samples, and surfaces are fitted results*

## 4.4 Complex number regression

We give the complex number regression example with a complex number logarithm. The natural logarithm for the complex number can be analytically written as,

$$\ln(z) = \ln|z| + i\arg(z)$$

where $|z|$ is the complex modulus and $\arg(z)$ is the complex argument., The complex logarithm is the conformal map with a branch cut in the complex plane $(-\infty, 0]$. Here, in figure 5, we give the estimate of $\ln|z|$ on equidistance data. Figure 5-a is the theoretical value of $|\ln(z)|$, b) is the values of $|\ln(z)| + \varepsilon$, and c) is the kernel estimate $|\ln(z)|$ using the code of FKreg with LL regression. The uppers are 2d top-down viewing of lowers. We then apply the LL regression with the same example but with uniform space shown in Figure 6, a)-d) are the $\text{Real}(\ln(z))$, $\text{Imag}(\ln(z))$, $|\ln(z)|$ and $\text{angle}(\ln(z))$. The blue dots are the data fed into the regression function.

```
x=[2*rand(n^2,1)+0.1,4*rand(n^2,1)-2];
x=x(:,1)+1i*x(:,2);
y=log(x);
yr=y+0.1*std(real(y)).*randn(size(y))+0.1*1i*std(imag(y)).*randn(size(y));
hlist=get_hList(20,[0.1,1],@logspace);
opt.order=1;
opt.dstd=3;
[regs]=cv_fkreg(x,yr,hlist,opt);
yhat=fkreg_predict(regs,x);
```

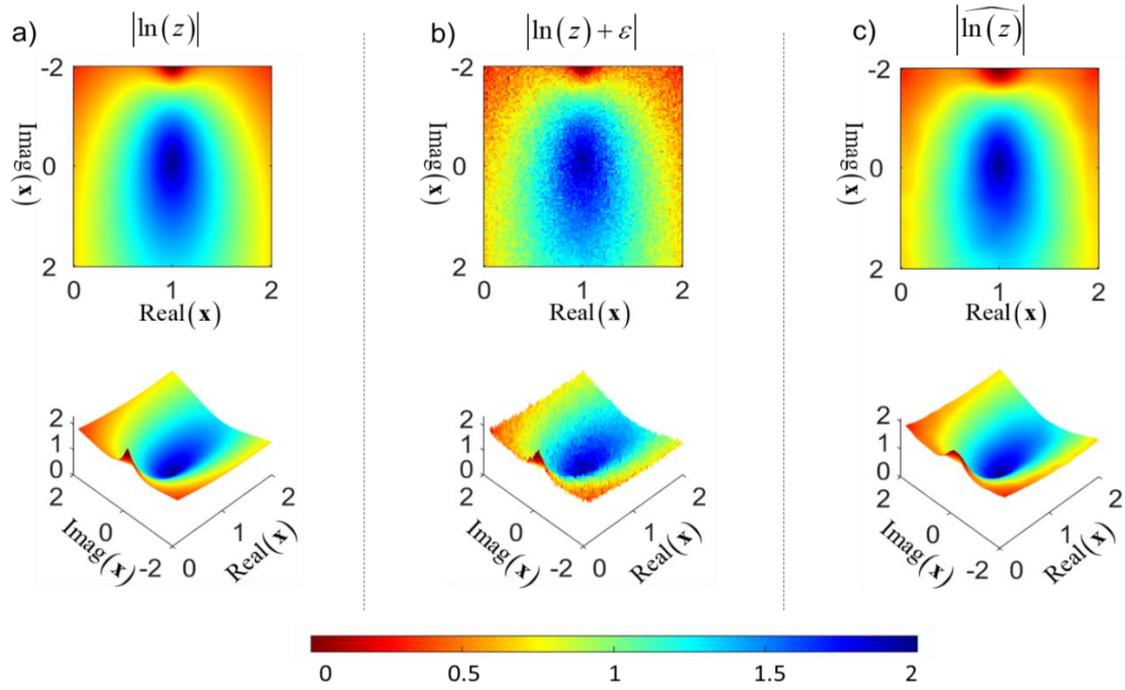

Figure 8: Example of complex value regression. a) the magnitude of $\ln(z)$; b) the magnitude of *observation* $\ln(z)+\varepsilon$ ; c) the magnitude of $\ln(z)$.

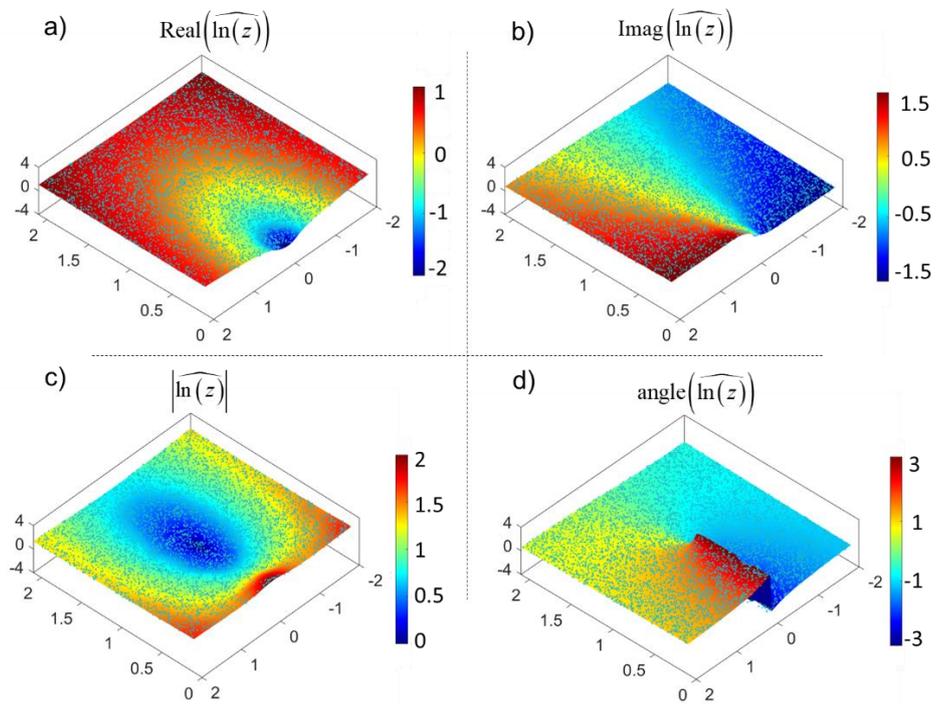

Figure 9: The result of complex value regression. a) the real value of $\ln(z)$; b) the imaginary value of $\ln(z)$ ;c) the magnitude of $\ln(z)$; d) the angle of $\ln(z)$

## 4.5 Application on EEG developmental surface

Except for the simulation, we apply this routine to the real quantitive EEG (qEEG) study (the dataset can be downloaded from the web: https://doi.org/10.7303/syn26712979). This application aims to show the developmental surface of the human brain from EEG, which is the mean trajectory of descriptive parameters (DPs) with frequency and age. For this example, use the EEG log-spectrum $y_{s,\omega,c}$ of the subject $s$ at the frequency $\omega$ for the electrode $c$ as DPs, where $s=1,\ldots,Ns$ $\omega=\Delta\omega,\ldots,N\omega\Delta\omega$ and $c=1,\ldots,Nc$. Thus, the multip-input observation is a matrix $Y$ with the size $(N\omega \cdot Ns) \times Nc$ that the columns are independent log-spectrum vectors for all electrodes and share the same smooth space composed of all subjects' age and the log spectrum's frequency points.

Here, the size of the input $Y$ is $47 \times 1564 \times 18$. To successfully execute large-multiple regression, in the function **fkreg_y,** we have separated $Y$ to save memory according to the memory cost estimation. Figure 8 takes the developmental surface of the left occipital electrode (O1) as an example. Figure 6-a is the mean log-spectrum surface with frequency and age with the raw scatter plot in subplot b. The result shows that the alpha peak of the log-spectrum moves forward from child to young people and moves back when the age increases, which can also be seen on the top view in subplot b. The result is consistent with the conclusion of Szava et al. (1994) and has been used in Li et al. (2022).

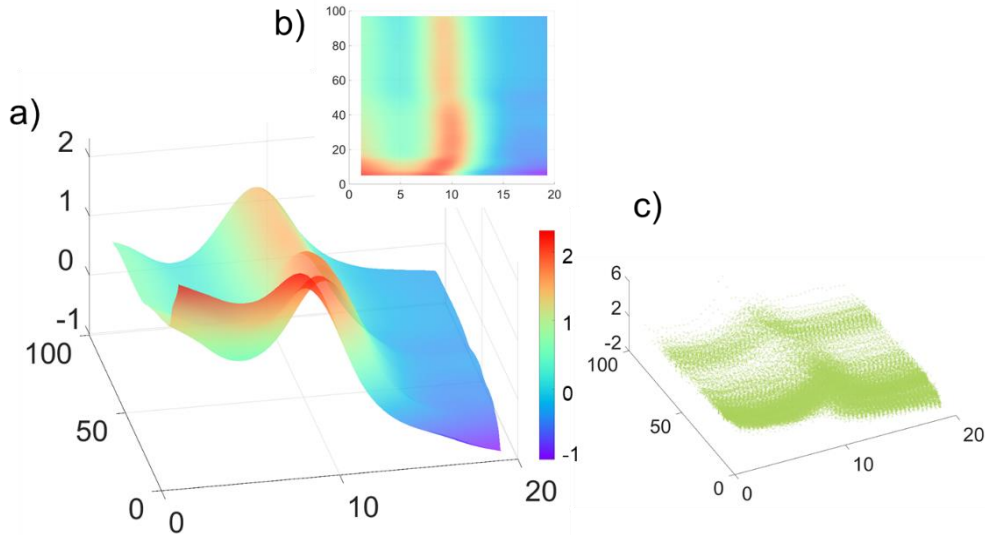

Figure 10: The Brain developmental surface. a) The shape of EEG log-spectral with frequency and age, which is the mean value of sample log-spectral values shown in c); b) The top view of a); c) sample log-spectral values scatter plot

## 5 Speed Test

Under the condition of $N \ll M$, the complexity of the regression method proposed in this paper is mainly contributed by NUFFT, especially its fast gaussian gridding, which has $O(N)$ complexity.

We compared our regression procedure with the fast sum updating toolbox (Gevret et al., 2018; Langrené and Warin, 2019) for 1 dimension data. Both methods used a fixed bandwidth without calculating the DF. In order to handle a large enough sample size, the NUFFT-regression ran on the configuration of 100 grids points with 1 digital accuracy and the prediction method of griddedinterpolant set with linear both for interpolation and extrapolation. The datasets were randomly sampled from 10^1 to 10^9 points, and the experiment was executed on an AMD Ryzen 7 3700X CPU PC with 32×4GB 3600Mhz memory.

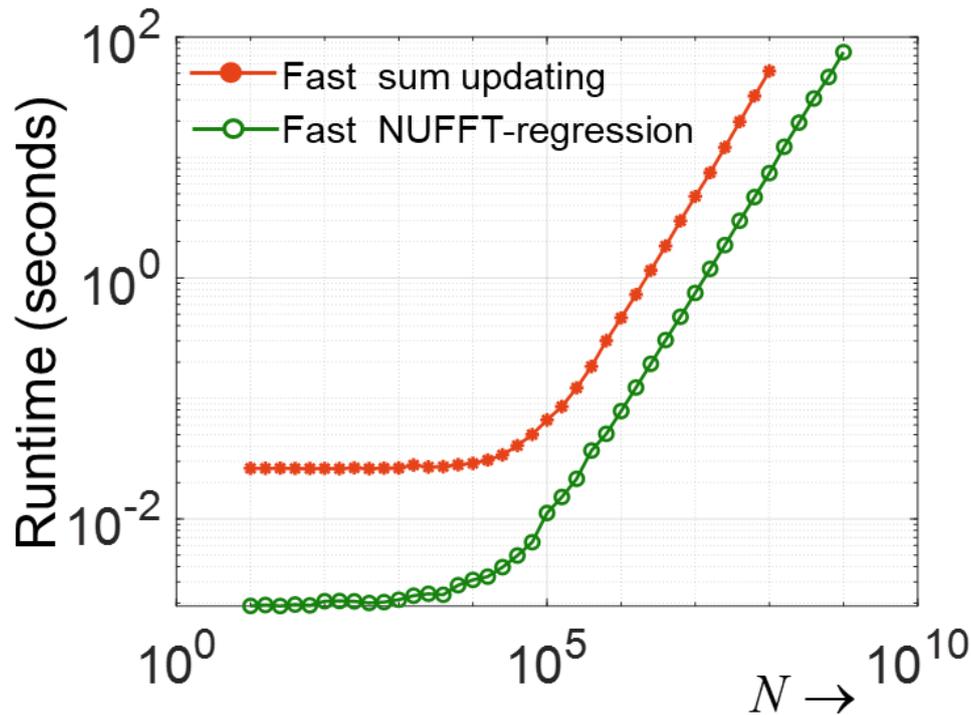

*Figure 11 Runtime for both toolbox with 10^0.2 interval, results take median elapsed time of 10 times independent evaluation.*

Figure 11 shows that our toolbox of this paper can carry out large datasets. The execution time for a single smooth is efficient, for 10^9 points can be finished in less than 100 seconds. For a typical size dataset (<10^7 samples), the time elapsed was less than 1 second for a single smooth.

## 6  Discussion

The complexity of kernel regression is a big challenge for nonparametric regression, especially for non-uniform sampled data. It limits its application to the large sample size data and real-time system. The fast-binning strategy offers a simple but effective way to solve it. Unfortunately, there's no toolbox available. Besides providing the toolbox, this paper improves its accuracy, coordinates with a fast bandwidth selection method, and extends it to multiple and multivariate cases. Additionally, the typical application for the variance estimation of the heteroscedasticity model is also included in the toolbox.

The algorithm of the fast kernel regression with NUFFT is basically composed of two steps gridding and fast convolution. The gridding convolves the discrete samples with a broad bandwidth kernel and transforms the result into the spectral domain. In the spectral domain, we deconvolve the effect of the gridding kernel to ensure the frequency transform equals the DFT of the original data. Then the transformed data will be applied with the local polynomial regression in the frequency domain, which directly element-wise products the kernel function and the data. Finally, we can transform it back to the raw space and use a low-cost interpolation method on the gridded data.

Flowing the idea of simplicity and efficiency, the implementation of the toolbox was in pure MATLAB code and took advantage of the efficiency of its vectorized operations. The multivariate high order local polynomial can be solved with a clear symbolic calculation to get the estimator. There are still some trivia issues for fast-binning regression that cannot be ignored.

- High order local polynomial regression might have the stabile problem because of the inversion of the design matrix, which might be solved by adding the constrain term. Anyhow, the local linear is suggested. As Fan and Gijbels, (1992) said, it is conditional unbiased. The result in figure 4 also shows it keeps the local variability pattern and evaluates fast.
- How many bins should take in each direction? Usually, the choice of the number of bins is much smaller than the number of points in the original space, but there are no precise settings. The number of bins in each direction does not change very much between dimensions (Wand, 1994). Definitely, in some applications, the number of bins can be larger than the input for high resolution.
- What is the number of the spread kernel bandwidth for binning? The value of $\tau$ dependent on the accurate wished. For example, $\tau = 12/N^2$ the gaussian spreading of each source to the nearest 24 grid points (12 points for each side) yields about 12 digits of accuracy (Alok Dutt,1993). Instead of the linear and simple binning, which Wand compared these two methods in 1994 (Wand, 1994) that the linear binning method convergent faster than simple binning, the gaussian gridding method can achieve higher accuracy by setting the number of accuracies and computing the same as the linear binning method. Furthermore, the application of fast Gaussian gridding will add brilliance to gaussian gridding's present splendor.
- How to select the kernel? The class of kernel bandwidth selection has been discussed a lot, see (Čížek and Sadıkoğlu, 2020; Langrené and Warin, 2019) in depth.

- The bandwidth is revealed on the cutoff frequency, which might be used for designing a better selection method based on the distribution of data in its spectral domain.
- The multi-response regression in this toolbox does not take into account the influence between variables.
- The robust regression, including the missing data problem and outlier regression, hasn't been discussed here, which will be the next step for the toolbox.

Our FKreg toolbox is designed easy to start, providing efficient codes for nonparametric function estimation. The toolbox can be used in the normative study(Dinga et al., 2021; Li et al., 2022; Marquand et al., 2019). The fast local polynomial regression with the rGCV method for multiple multivariate estimations can efficiently handle such big datasets.